# Investigating the Effect of Display Refresh Rate on First-Person Shooting Games

Haoshen Qin, Zixian Zhu
h.qin@ufl.edu, zhuzixian@ufl.edu
University of Florida, Gainesville, FL, USA

## Abstract

For first-person shooting game players, display refresh rate is important for a smooth experience. Multiple studies have shown that a low display refresh rate will reduce gamers' experience and performance. However, the human eye's perception of refresh rate has an upper limit, which is usually less than what high-performance monitors, for which players pay much higher prices, provide. This study assesses whether a higher refresh rate always has a positive impact on players' performance, making it worthwhile for them to invest in high-performance monitors. A within-group experimental design study was conducted using a commercial first-person shooting game platform (N = 26) to investigate players' performance at display refresh rates of 30Hz, 60Hz, 120Hz, 144Hz, and 240Hz. Player performance was assessed based on score, accuracy, and self-ratings from the players. The results show that display refresh rate only significantly affects player performance at 30Hz.

## Introduction

Video games are one of the most popular entertainment ways of the world, and First-person shooting (FPS) game is a dominant genre within the video game industry. As of 2023, there are 4 FPS games in the top 10 most-played games. In that list, the most played game is Player's Unknow Battlegrounds, which is a FPS game. (Naeem, 2024) The increasing popularity of video games has correspondingly increased the demand for high performance gaming hardware. The game player often attempts to intensify the immersive world of video games with newer versions of hardware that are made for the newest technology, such as graphics cards, CPU, and monitor with high refresh rates. This equipment can improve player's game experience by providing smoother visuals and quicker response times, which are critical in the game you need quick response times like FPS game. There are some studies already shown the effects of display refresh rate on FPS games (Kajal & Mark, 2007). However, the previous study only focusses refresh rate around 144 hz range (Huhti, 2019) and the targeted group focus on high-skilled players (Spjut et al., 2019). The high-performance monitors can now provide over 300hz refresh rate, and the benefits of high-performance monitors, especially those that offer extremely high refresh rate, have yet to be systemically proven, especially for general players rather than professional well-trained players.

Our study aims to determine whether a higher display refresh rate benefits common players who are not well-trained professionals but still seek improved game performance with more expensive devices. We investigated this research question through an empirical experiment to gain insights into the effect of higher refresh rates on common players' first-person shooting performance. This research is crucial for providing advice to common players on whether expensive hardware upgrades truly enhance their FPS game experience.

## Methodology

The hypothesis of the study can be concluded as follows:

    H₀: The refresh rate does not significantly influence the player's performance in FPS games.

    Hₐ: The refresh rate significantly influences the player's performance in FPS games.

A within-subject experiment (N = 26) was conducted to assess the hypothesis (UF IRB Exempted #ET00022986),

considering participants were recruited with varying backgrounds in gaming experience and skills, and the individual differences in their gaming performance could be substantial. Each participant needed to complete the same one-minute shooting task from a commercial FPS gaming platform during the experiment under five different conditions with refresh rates set to 30Hz, 60Hz, 120Hz, 144Hz, and 240Hz, which are all commonly used default computer monitor refresh rate. The sequence of conditions they experienced was randomly assigned, and they received at least three practice sessions to achieve stable task performance before experiencing the conditions. Participants were only informed to perform the FPS game under 5 conditions but were not provided with any information regarding the refresh rate or other experimental settings to avoid any bias or psychological impacts. The study's purpose will be disclosed to the participants after they have completed all conditions.

**Study Measurements** The player's performance will be assessed based on Score, Accuracy, and Self-rating. The score and accuracy were directly obtained from the commercial gaming platform, representing the objective performance of each condition (Figure 1). The score is the cumulative score based on the number of targets hit by the participants, the precision of the target locations hit, and the targets missed during shooting. Accuracy is the percentage of overall targets hit by the participants during shooting. Self-rating is a 5-point Likert scale that allows participants to indicate how successful they felt during the task. Participants also have the option to provide comments if they did not feel successful.

**Equipment and Settings** The study provided identical environments and devices for each participant. The gaming device used was an Alienware Laptop with an Intel® Core™ i9-13980HX Processor, GeForce RTX 4060 Graphics Card, and a 1920 x 1080 resolution FHD monitor with a maximum refresh rate of 240Hz. The participants also performed the task in the same location, using the same seat, mouse, and mouse pad (Figure 2). The participants were provided with sufficient rest time during the experiment until they felt ready to begin the game. A pilot study was conducted to ensure that all experimental settings and procedures functioned properly for data collection.

**Experiment Procedure** The participants first accessed the online survey by scanning the QR code. After signing the experiment consent form, they filled out demographic surveys about gaming frequency, gaming skill, and gaming genres. The participants were then provided with three practice sessions for their experimental task to achieve stable performance and mitigate bias from learning curves. If the participants' gaming scores kept increasing during the three practice sessions, they were provided with more practice sessions until the scores dropped or remained the same. During the practice sessions, the refresh rate was set to 240Hz. The participants then experienced the 5 conditions with different refresh rate settings (Condition 1 - 30 Hz, Condition 2 - 60 Hz, Condition 3 - 120 Hz, Condition 4 - 144 Hz, and Condition 5 - 240 Hz), based on their randomly assigned sequence. They were informed that they could rest in another location within the experiment room and could not see the setting changes before each condition started. After completing the one-minute task, they were able to fill out self-rating surveys regarding their performance in that condition and had substantial rest time until they were ready for the next condition.

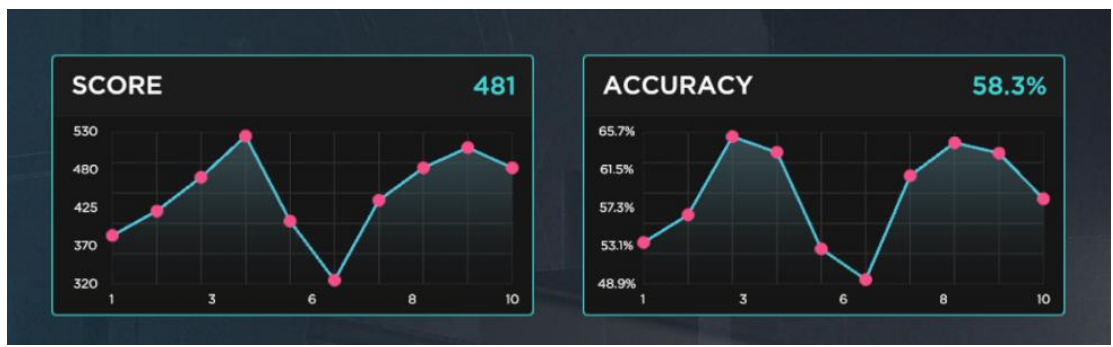

Figure 1. Score and accuracy of one-minute shooting task

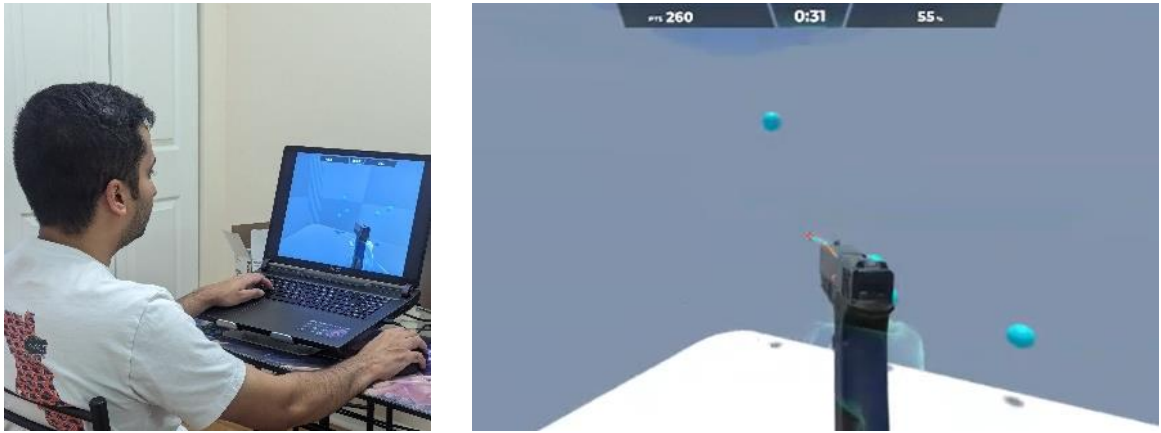

Figure 2. Experiment environment and the player's view during task

## Results

The age of the 26 participants ranged from 22 to 31 (Average = 26.19). 15 participants are male, 10 participants are female, and one participant did not report their gender. Based on the question regarding gaming genres, 8 participants reported that they enjoy playing FPS games in their free time. The distribution of gaming frequency and gaming skills are shown in Figure 3, indicating that the distribution decreases as the gaming skill level increases. The distribution of favorite game genres among participants shows that action/adventure games are the most popular, with first-person shooter games and role-playing games ranking second.

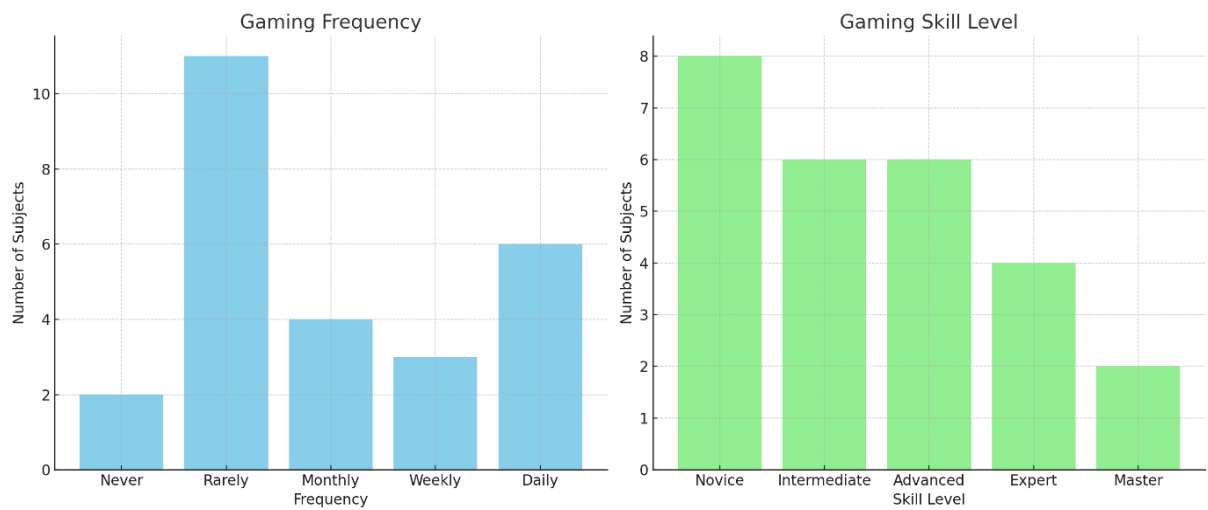

Figure 3. Distribution of participants' gaming frequency and skill level

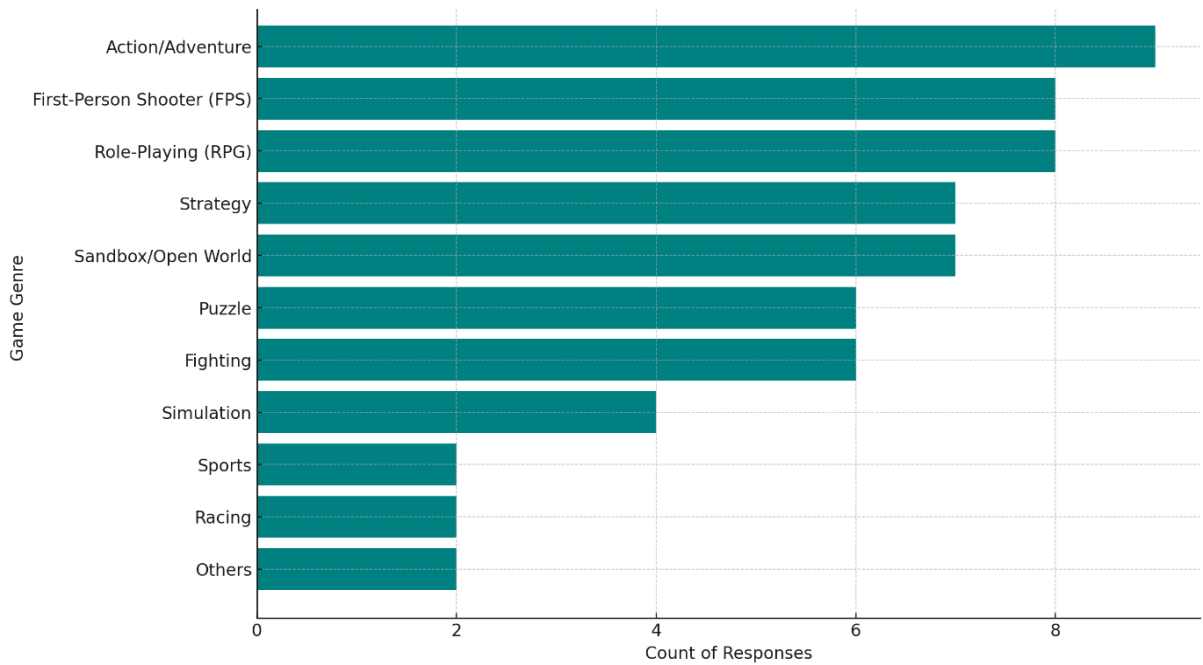

Figure 4. Popularity of game genres among participants

**Score** The distribution of the score data is showed in Figure 4. The results from Shapiro-Wilk test show that the Sig. value for condition 1 to 5 is all higher than 0.05 (0.448, 0.269, 0.635, 0.410, 0.333), which indicates that there is not significant evidence to conclude that the residuals are not sampled from a normal distribution. Therefore, a repeated ANOVA test was used to analyze if there is any significant influence from refresh rate for participants gaming score of all 5 conditions, which the refresh rate is 30hz, 60hz, 120hz, 144hz, and 240hz.

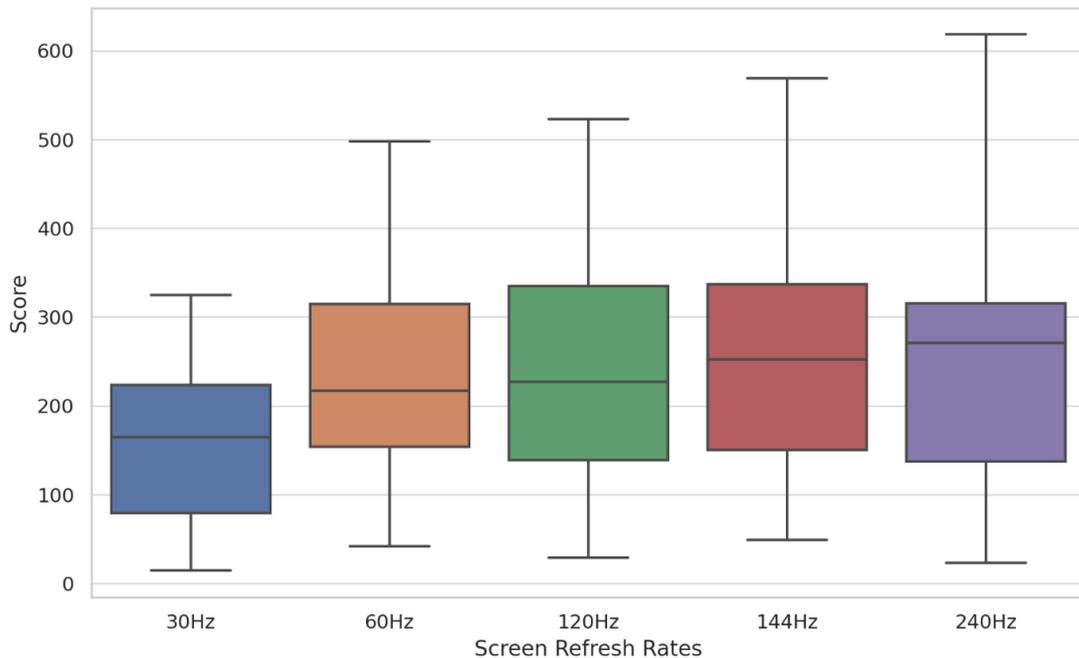

Figure 4. Distribution of Player's scores by refresh rate

The results show that players' scores were significantly affected by refresh rate, $F(2.76, 69.03) = 27.90$, $p < .001$.

Since Mauchley's test of sphericity was violated, the Greenhouse-Geisser correction was used. The effect size, as indicated by Eta$^2$ ($\eta^2 = 0.527$), suggested that the effect of fatigue on balance errors was substantial (see Table 1). Bonferroni post-hoc tests comparing refresh conditions revealed a significant difference only in players' scores between the 30 Hz refresh rate condition (condition 1) and all other conditions (p < .001). No other comparisons were significant among the conditions of 60 Hz, 120 Hz, 144 Hz, and 240 Hz (see Table 2).

Table 1. Test of within-subject effects of refresh rate on players' scores

| Source | Sum of Squares | df | Mean Square | F | Sig. | Partial Eta Squared |
|---|---|---|---|---|---|---|
| **Refresh rate** | 212977.9 | 2.761 | 77132.52 | 27.902 | <.001 | 0.527 |
| **Error** | 190826.1 | 69.03 | 2764.399 | | | |

Table 2. Pairwise Comparison with Bonferroni Adjustments for the effects of refresh rate on players' scores

| (I) condition | (J) condition | Mean Difference (I-J) | Std. Error | Sig.$^b$ |
|---|---|---|---|---|
| 1 | 2 | -88.385* | 11.643 | <.001 |
|   | 3 | -93.731* | 13.655 | <.001 |
|   | 4 | -107.962* | 13.867 | <.001 |
|   | 5 | -107.423* | 16.783 | <.001 |
| 2 | 1 | 88.385* | 11.643 | <.001 |
|   | 3 | -5.346 | 10.115 | 1 |
|   | 4 | -19.577 | 10.341 | 0.7 |
|   | 5 | -19.038 | 13.414 | 1 |
| 3 | 1 | 93.731* | 13.655 | <.001 |
|   | 2 | 5.346 | 10.115 | 1 |
|   | 4 | -14.231 | 8.008 | 0.877 |
|   | 5 | -13.692 | 10.134 | 1 |
| 4 | 1 | 107.962* | 13.867 | <.001 |
|   | 2 | 19.577 | 10.341 | 0.7 |
|   | 3 | 14.231 | 8.008 | 0.877 |
|   | 5 | 0.538 | 10.767 | 1 |
| 5 | 1 | 107.423* | 16.783 | <.001 |
|   | 2 | 19.038 | 13.414 | 1 |
|   | 3 | 13.692 | 10.134 | 1 |
|   | 4 | -0.538 | 10.767 | 1 |

Based on estimated marginal means

* The mean difference is significant at the .05 level.

b Adjustment for multiple comparisons: Bonferroni.

**Accuracy** The distribution of the accuracy (%) is showed in Figure 5. The results from Shapiro-Wilk test show that the Sig. value for condition 1 to 5 is all higher than 0.05 (0.690, 0.455, 0.102, 0.653, 0.883), which indicates that there is not significant evidence to conclude that the residuals are not sampled from a normal distribution. Therefore, a repeated ANOVA test was used to analyze if there is any significant influence from refresh rate for participants accuracy of all 5 conditions. Mauchley's test of sphericity was not violated for accuracy data. The results show that players' scores were significantly affected by refresh rate, $F(4, 100) = 13.776$, $p < .001$ (Table

3). Bonferroni post-hoc tests comparing refresh conditions revealed a significant difference only in players' accuracy between the 30 Hz refresh rate condition (condition 1) and all other conditions (p < .001). No other comparisons were significant among the conditions of 60 Hz, 120 Hz, 144 Hz, and 240 Hz (see Table 4).

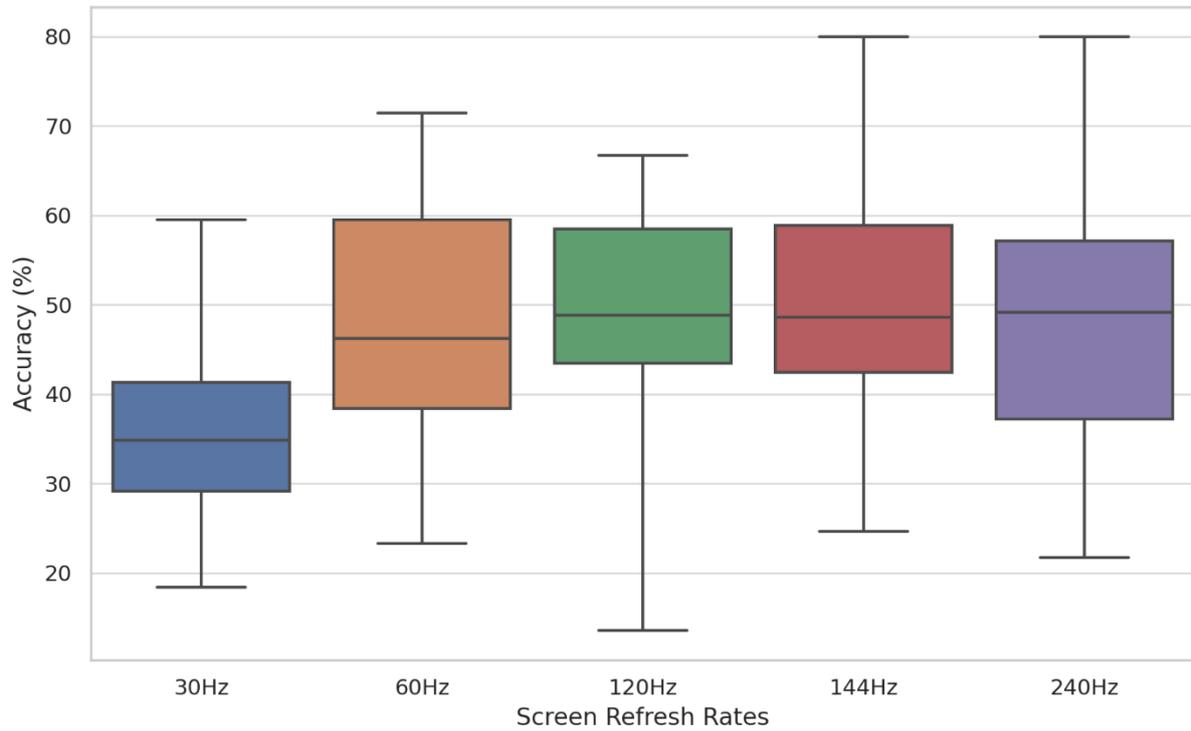

Figure 5. Distribution of player's accuracy by refresh rate

Table 3. Test of within-subject effects of refresh rate on players' accuracy

| Source | Sum of Squares | df | Mean Square | F | Sig. |
|---|---|---|---|---|---|
| **Refresh rate** | 3858.35 | 4 | 964.588 | 13.776 | <.001 |
| **Error** | 7001.854 | 100 | 70.019 | | |

Table 4. Pairwise Comparison with Bonferroni Adjustments for the effects of refresh rate on players' accuracy

| (I) condition | (J) condition | Mean Difference (I-J) | Std. Error | Sig.[b] |
|---|---|---|---|---|
| 1 | 2 | -12.400* | 2.028 | <.001 |
|   | 3 | -13.423* | 2.371 | <.001 |
|   | 4 | -15.008* | 2.694 | <.001 |
|   | 5 | -12.950* | 2.605 | <.001 |
| 2 | 1 | 12.400* | 2.028 | <.001 |
|   | 3 | -1.023 | 2.057 | 1 |
|   | 4 | -2.608 | 2.024 | 1 |
|   | 5 | -0.55 | 2.344 | 1 |
| 3 | 1 | 13.423* | 2.371 | <.001 |
|   | 2 | 1.023 | 2.057 | 1 |
|   | 4 | -1.585 | 1.513 | 1 |
|   | 5 | 0.473 | 2.584 | 1 |

| | | | | | |
|---|---|---|---|---|---|
| 4 | 1 | 15.008* | 2.694 | <.001 | |
| | 2 | 2.608 | 2.024 | 1 | |
| | 3 | 1.585 | 1.513 | 1 | |
| | 5 | 2.058 | 2.7 | 1 | |
| 5 | 1 | 12.950* | 2.605 | <.001 | |
| | 2 | 0.55 | 2.344 | 1 | |
| | 3 | -0.473 | 2.584 | 1 | |
| | 4 | -2.058 | 2.7 | 1 | |

Based on estimated marginal means

\* The mean difference is significant at the .05 level.

b Adjustment for multiple comparisons: Bonferroni.

**Self-rating** The distribution of the 5-level Likert scale self-rating is showed in Figure 6. The results from Shapiro-Wilk test show that the Sig. value for condition 1 to 5 is all less than 0.05, which indicates that there is significant evidence to conclude that the residuals are not sampled from a normal distribution. Therefore, a Friedman's Two-Way ANOVA test was used to analyze if there is any significant influence from refresh rate for participants self-rating of all 5 conditions. The results show that players' self-rating was significantly affected by refresh rate, $p < .001$. Bonferroni post-hoc tests comparing refresh conditions revealed a significant difference only in players' self-rating between the 30 Hz refresh rate condition (condition 1) and all other conditions ($p < .001$). No other comparisons were significant among the conditions of 60 Hz, 120 Hz, 144 Hz, and 240 Hz (Table 5, Figure 7).

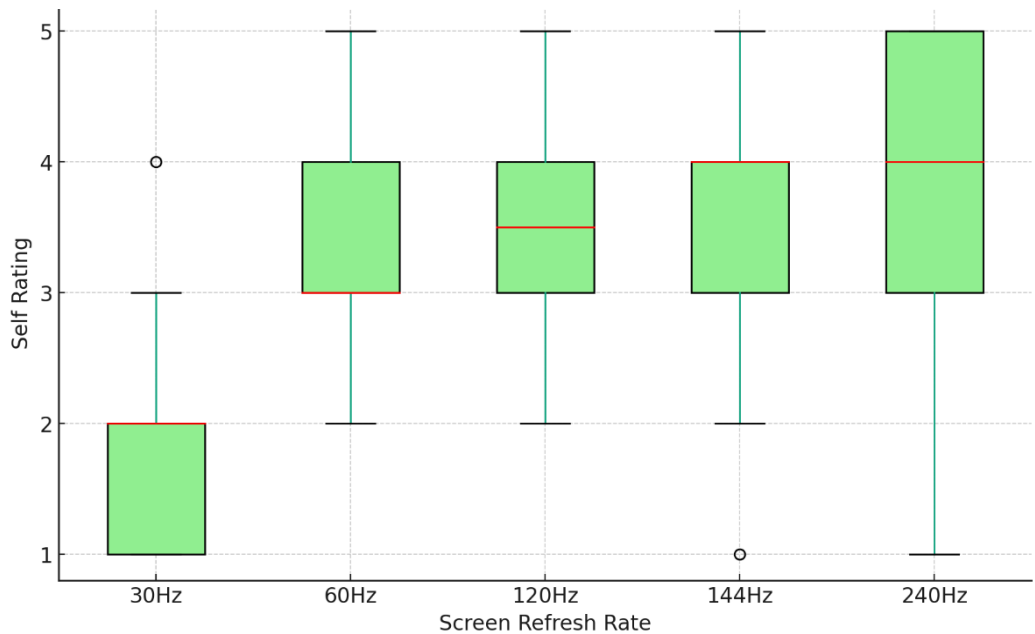

Figure 5. Distribution of player's self-rating by refresh rate

Table 5. Pairwise Comparison with Bonferroni Adjustments for the effects of refresh rate on players' self-rating

| Sample 1-Sample 2 | Test Statistic | Std. Error | Std. Test Statistic | Sig. | Adj. Sig.[a] |
|---|---|---|---|---|---|
| condition1-condition3 | -2 | 0.439 | -4.561 | <.001 | 0 |
| condition1-condition2 | -2.058 | 0.439 | -4.692 | <.001 | 0 |
| condition1-condition4 | -2.154 | 0.439 | -4.912 | <.001 | 0 |
| condition1-condition5 | -2.346 | 0.439 | -5.35 | <.001 | 0 |
| condition3-condition2 | 0.058 | 0.439 | 0.132 | 0.895 | 1 |
| condition3-condition4 | -0.154 | 0.439 | -0.351 | 0.726 | 1 |
| condition3-condition5 | -0.346 | 0.439 | -0.789 | 0.43 | 1 |
| condition2-condition4 | -0.096 | 0.439 | -0.219 | 0.826 | 1 |
| condition2-condition5 | -0.288 | 0.439 | -0.658 | 0.511 | 1 |
| condition4-condition5 | -0.192 | 0.439 | -0.439 | 0.661 | 1 |

Each row tests the null hypothesis that the Sample 1 and Sample 2 distributions are the same.

Asymptotic significances (2-sided tests) are displayed. The significance level is .050.

Figure 7: Heatmap of pairwise comparison with Bonferroni adjustment for refresh rate effects on player's self-ratings

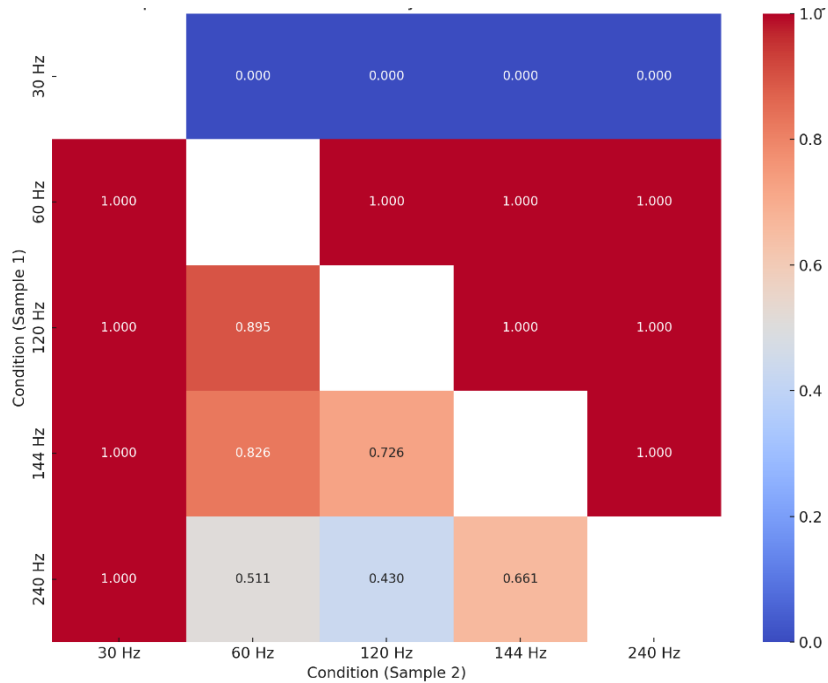

In self-rating comments, 7 out of 26 participants indicated that they felt the refresh rate or the FPS (frames per second) was low when they were doing condition 1 (30Hz refresh rate). 4 of them reported the self-rating task success level as 2, and 3 participants reported success level 1. 2 of the participants also reported that they felt the FPS or the refresh rate improved in the next condition after condition 1 (30Hz refresh rate). Among these 7 participants who noticed the independent variable for the conditions, 6 out of 7 reported in the demographic survey that they enjoyed First-Person Shooter (FPS) games the most among all game genres. There were other comments on condition 1 reporting that they felt "it's the most difficult one," "I think the game was lagging a little this time," and "I felt the targets moved slower," indicating that the players felt a difference in condition 1 compared to other conditions. Comments on other conditions were only regarding their own performance rather than the perception of the system, such as "I missed many targets," "I had better accuracy than last time," and "I was too nervous."

Considering there is no significant difference in score, accuracy, and self-rating between the 60Hz, 120Hz, 144Hz, and 240Hz conditions, a power analysis was conducted using G*Power software. Based on Cohen's d, with an effect size of 0.25 (indicating a small effect size) and a power of 0.8, the required sample size is 21, which is smaller than our study's sample size of 26, indicating that the results can provide statistical insights.

**Conclusion and Discussion**

The results of this study show that there is a significant effect of monitor refresh rates on player performance in FPS games. And refresh rates have a crucial role in both objective measures (scores and accuracy) and subjective measures (such as self-rated). There is significant difference between the lowest refresh(30HZ) rate and higher refresh rate (60HZ – 240HZ). It also suggests that the fluidity and smoothness provided by higher refresh rates can enhance the player's experience and performance. However, the lack of significant differences between higher refresh rates(120HZ, 144HZ, 240HZ) may indicate diminishing returns from perceived improvements beyond a specific threshold. This is consistent with the findings of related studies, which show that above 144HZ, most gamers may be difficult to detect performance improvements (Hagström, September, 2015 ). While the most significant performance enhancement was observed when moving from 30HZ to higher refresh rates, this advantage appears to stable at refresh rates higher than 144HZ. This study also highlights the sensitivity of experienced FPS game players to changes in refresh rates, as evidenced by their comments and self-ratings. This suggests that the ability to discern differences in refresh rates may be more pronounced among individuals who are more familiar with the dynamics of these games, which may influence their preferences and perceptions of game quality. This finding is particularly important for players considering hardware upgrades and the manufacturers targeting this consumer group. Players who are frequently more involved in FPS gaming or who are particularly sensitive to visual fluidity may find this investment justified. Instead, casual players can choose a moderately to high refresh rate (60Hz – 120Hz) monitor, so they can still enjoy improved performance without paying the high costs associated with monitors that have high refresh rates.

**Acknowledgement**

We would like to thank all the participants who invested their time and effort in this study. Special thanks to Tongyang Wang for his support in equipment.

**Reference**

Ipcstore.com. The most played games of 2023. Online: https://www.ipcstore.com/blog/the-most-played-games-of-2023.

Claypool, K.T., Claypool, M. On frame rate and player performance in first person shooter games. Multimedia Systems 13, 3–17 (2007). https://doi.org/10.1007/s00530-007-0081-1

Hagström, R. (2015). Frames That Matter : The Importance of Frames per Second in Games (Dissertation). Retrieved from https://urn.kb.se/resolve?urn=urn:nbn:se:uu:diva-263379

Spjut, J., Boudaoud, B., Binaee, K., Kim, J., Majercik, A., McGuire, M., Luebke, D., & Kim, J. (2019). Latency of 30 ms Benefits First Person Targeting Tasks More Than Refresh Rate Above 60 Hz. SIGGRAPH Asia 2019 Technical Briefs, 110–113. https://doi.org/10.1145/3355088.3365170

Huhti, J. P. (2019). The Effect of High Monitor Refresh Rate on Game Experience. https://osuva.uwasa.fi/handle/10024/10196